# Hourly Warning for Strong Earthquakes

T. Chen[1*], L. Li[1,2], X.-X. Zhang[3], C. Wang[1] (Chi Wang), X.-B. Jin[4], Q.-M. Ma[5], J.-Y. Xu[1], Z.-H. He[1], H. Li[1], S.-G. Xiao[1], X.-Z. Wang[6], X.-H. Shen[7], X.-M. Zhang[8], H.-B. Li[9], Z.-M. Zeren[7], J.-P. Huang[7], F.-Q. Huang[10], S. Che[8], Z.-M. Zou[1], P. Xiong[8], J. Liu[8], L.-Q. Zhang[1], Q. Guo[7], I. Roth[11], V. S. Makhmutov[12], Yong C.-M. Liu[1] (Yong Liu), Z.-H. Huang[1], J. Luo[1], X.-J. Jiang[1], L. Dai[1], S.-P. Duan[1], X.-D. Peng[1], X. Hu[1], H. Wu[1], S. Ti[1], C. Zeng[1], J.-J. Song[5], F. Xiao[5], J.-G. Guo[3], C. Wang[3] (Cong Wang), L. Yao[10], A.-M. Du[13], Y. Wei[13], H. Yuan[14], S. Wang[14], H.-Y. Cui[15], C. Li[15], Y. Liu[16] (Yi Liu), J. Yang[16], Q. Sun[16], J.-F. Su[1], W. Li[1], Y.-C. Zhang[1], D.-L. Li[1], S.-H. Wang[1,2], C.-L. Cai[1], G.-Q. Yan[1]

[1]State Key Laboratory of Space Weather, National Space Science Center, Chinese Academy of Sciences, Beijing, China.

[2]College of Earth and Planetary Sciences, University of Chinese Academy of Sciences, Beijing, China.

[3]National Center for Space Weather, China Meteorological Administration, Beijing, China.

[4]Sichuan Meteorological Disaster Defense Center, Chengdu, China.

[5]Institute of Electrical Engineering, Chinese Academy of Sciences, Beijing, China.

[6]Institute of Geophysics, China Earthquake Administration, Beijing, China.

[7]National Institute of Natural Hazards, Ministry of Emergency Management of China, Beijing, China.

[8]Institute of Earthquake Forecasting, China Earthquake Administration, Beijing, China.

[9]Institute of Geology, Chinese Academy of Geological Sciences, Beijing, China.

[10]China Earthquake Networks Center, China Earthquake Administration, Beijing, China.

[11]Space Sciences Laboratory, UC Berkeley, CA, USA.

[12]Lebedev Physical Institute, Russian Academy of Sciences, Moscow, Russia.

[13]Institute of Geology and Geophysics, Chinese Academy of Sciences, Beijing, China.

[14]Academy of Opto-Electronics, Chinese Academy of Sciences, Beijing, China.

[15]Institute of Acoustics, Chinese Academy of Sciences, Beijing, China.

[16]Institute of Atmospheric Physics, Chinese Academy of Sciences, Beijing, China.

*Corresponding author. Email: tchen@nssc.ac.cn

**Abstract:** A promising perspective is presented that humans can provide hourly warning for strong land earthquakes (EQs, Ms≥6). Two important atmospheric electrostatic signal features are described. A table that lists 9 strong land EQs with shock time, epicenter, magnitude, weather in the region near the epicenter, precursor beginning time, and precursor duration demonstrates that at approximately several hours to one day before a strong land EQ, the weather conditions are fair near the epicenter, and an abnormal negative atmospheric electrostatic signal is very obvious. Moreover, the mechanism is explained. A method by which someone could determine the epicenter and the magnitude of a forthcoming strong EQ is suggested. Finally, the possibility of realizing hourly warning for strong land EQs in the near future is pointed out.

**One-Sentence Summary:**

A method that provides the information of the shock time, the epicenter and the magnitude of a strong land EQ is suggested.

**Main Text:**

Due to the devastating consequences of strong earthquakes, the identification of any reliable precursor signatures is of paramount importance. There are many kinds of parameter anomalies that could be applied as earthquake precursors (*1-3*): Geostress changes; underground water level and other fluid changes; the geoelectric field, geoconductivity and geomagnetic disturbance records; gravitational anomalies; surface deformation; very large gas fluxes out of the crust; variations in radon gas; temperature variations at the Earth's surface; air temperature variations; variations in air relative humidity; an anomalous flux of latent heat of evaporation; extraordinary vertical profiles of air temperature and humidity; linear cloud anomalies; anomalies of radio wave propagation in very low frequency (VLF), high frequency (HF) and very high frequency (VHF) bands; extraordinary concentrations and distributions of aerosols; anomalies of the outgoing longwave radiation (OLR) energy flux; local (in situ) anomalies of space plasma parameters (concentrations of ions and electrons, ion and electron temperatures, mass compositions and concentrations of the major ions); extraordinary extremely low frequency (ELF) and VLF emissions measured onboard a satellite; quasi-constant magnetic and electric fields; extraordinary particle precipitation fluxes for different energy bands; vertical profiles of the electron concentration; and extraordinary total electron content (TEC) change by GPS data processing (*1*). In addition, atmospheric electrostatic field monitoring has recently become more appealing. The atmospheric electric field anomalies prior to earthquakes have been widely studied (*4-11*). Since the Tangshan earthquake in 1976, China has set up several monitoring stations of atmospheric electric fields, and some obvious anomalous cases have been observed (5, *10*). Omori et al. (*11*) pointed out that anomalous radon emissions trigger great increases in the number density of small (or light) ions and the atmospheric conductivity and a decrease in the atmospheric field of the lower atmosphere (from the ground to an altitude of 2 km), as observed around the time of the Kobe earthquake in 1995. Omori et al. (*11*) further suggested that the behavior of radon in terms of the atmospheric electrical quasistatic process can explain seismic precursors observed near the ground. Choudhury et al. (*4*) described the characteristics of the vertical atmospheric electrostatic field as negative 7-12 hours before an earthquake according to the statistics of 30 earthquake events of various classes over northern India. Smirnov (*9*) reported that more than one hundred cases showed negative $E_z$ anomalies approximately one day in advance for Ms=4-6 earthquakes, but there is no obvious relation between the $E_z$ value and the epicenter, similar to $E_z$ and the magnitude. Altogether, a unique feature exists that the near-surface atmospheric quasistatic electric field appears anomalously negative just before some earthquakes with fair weather conditions (*10, 12*). This could be the most stable and reliable indicator that warns of an EQ to come in several hours to one day and alerts people of the emergency.

Although observations of the atmospheric electric field before an earthquake have been carried out before, they still cannot answer some outstanding questions: Where is the epicenter? What is the EQ magnitude? When does the EQ occur? On the one hand, it seems difficult to extract earthquake precursor information from atmospheric electric field anomalies because intense convection due to weather patterns, space weather effects, human activities, etc. may strongly interfere with the atmospheric ion background to produce ionization and change the motion of charged ions that are coupled to neutral molecules. On the other hand, there is a lack of sufficient and credible electric field data for researchers to study and improve the prediction of earthquakes by using negative atmospheric electric field anomalies (*5*). Additionally, until now, no compelling physical mechanism has existed that could explain why an abnormal "reverse" atmospheric electrostatic field signal could appear just before an earthquake. In this paper, we emphasize the importance of the anomalous negative signals of the atmospheric electric field to help predict some forthcoming earthquakes. First, we present two typical abnormal atmospheric electrostatic signal features that always appear several hours to one day before a strong land earthquake and a list of strong land earthquakes with shock time, epicenter, magnitude, weather condition, abnormal sign beginning time, and abnormal sign duration. Then, we offer an explanation of the electric field anomaly as a result of the enhanced production of ionization radiation due to the last tectonic activity preceding the earthquake. Finally, it is pointed that a monitoring network of the near-ground atmospheric electrostatic field anomalous negative $E_z$ signal may be very useful. Therefore, hourly warning of a strong land EQ is possible by using an $E_z$ monitoring network.

**Two Typical EQ Case Studies**

One typical feature (type 1) is shown in Fig. 1A. Despite the data gap, in Pixian and Wenjiang counties, $E_z$ acquired negative values for 7 hours before the Wenchuan Ms 8.0 earthquake, which occurred at 14:28 LT on May 12, 2008. Note that the weather in Wenchuan, Pixian and Wenjiang counties was very fair all day as well. The peak negative values of $E_z$ at Pixian and Wenjiang stations are -2750 V/m and -750 V/m, respectively. The Pixian station is 50 km from the epicenter, while the Wenjiang station is 55 km from the epicenter.

Another typical feature (type 2), as shown in Fig. 1B, was exemplified when the Changning Ms 6.0 earthquake occurred at 23:30 LT on June 17, 2019 (longitude=104.49°E, latitude=29.47°N) in Sichuan Province when at Yibin station situated 50 km from the epicenter, under fair air conditions, the $E_z$ became negative approximately 23 and a half hours before the EQ. The $E_z$ was suddenly reduced to a minimum of -10000 V/m at 23:30 LT (maximum of the negative $E_z$). The negative $E_z$ signal lasted approximately 70 minutes, and then Ez recovered to normal positive values. Twenty-three hours later, the Changning M6.0 earthquake occurred. This case is different from the first typical case in that $-E_z$ lasts only for a period, then returns to a positive $E_z$ state, and finally, the strong EQ occurs.

The above case studies demonstrated that there are two typical abnormal $-E_z$ signal characteristics: 1) Abnormal negative $E_z$ signals that appear and last until the occurrence of an impending earthquake, as shown in Fig. 1A; and 2) abnormal negative $E_z$ signals that last for a short time period and then return to the normal positive signals, and approximately several hours to one day later, the EQ occurs, as shown in Fig. 1B.

Table 1 shows the details of strong land earthquakes recorded in references, such as the three key elements of earthquakes (shock time, epicenter and magnitude) and characteristics of the electric field. All of these strong land earthquakes occurred on fair weather days, with negative electric field anomalies a few hours to a day before the EQ. Most of the strong earthquake abnormal negative $E_z$ signals lasted for a short time (approximately 1 hour). The values of the negative electric field signals range from a few hundred to tens of thousands of volts per meter, which may be related to the distance between the station and epicenter or the local geological structure. According to the two types of signals shown in Fig. 1, most strong land earthquakes are Type 2; that is, $E_z$ signals remained negative for a short time and then returned to positive until the earthquake occurred.

According to the above case studies and EQ list analysis, it is possible that if a strong land EQ is coming, first, the weather near the epicenter becomes fairer, and crustal movement near the related faults makes the near-surface atmospheric electrostatic field abnormal. The abnormal signal features can be divided into two types. One type is a negative $E_z$ signal lasting until the EQ shock. The other is a shorter negative period, with negative values generally maintained for only tens of minutes; then, the $E_z$ signal returns to normal. Additionally, other authors (*4, 9*) have stated that abnormal signals appear at the time scale of approximately several hours to one day. With these observation-based descriptions from all over the world, it is likely demonstrated that fair weather abnormal negative $E_z$ signals appearing several hours to one day before a strong land earthquake are universal phenomena.

**Why are Abnormal Negative Ez Signals so Close to the EQ Shock Time?**

Why does the atmospheric electrostatic field $E_z$ reversal last for several hours immediately before some earthquakes or until after an EQ? This could be explained by the fact that before the elastic rebound of the fault, crustal movement may cause the rock to enter a microfracture state, generating a large number of rock fractures and releasing a significant amount of radioactive radon gas into the atmosphere near the Earth's surface by these fractures.

The radon gases further decay and emit α particles, β particles and γ rays in the air. Compared to β particles and γ rays at various altitudes, α particles have a more powerful ionizing radiation capacity in air (*13*). Therefore, these α particles are the main source for ion pairs that include positively and negatively charged particles. In turn, regardless of the altitude at which they are produced, these positively charged particles could be driven downward to the Earth by the fair weather $E_z$ and gradually piled up above the surface. Furthermore, these negatively charged particles form a negatively charged layer at higher altitudes in the air. This gradient produces a reverse electric field. The magnitude of the reverse electric field may exceed that of the former downward $E_z$ in fair air (as shown in Fig. 2D).

According to the above observed cases, the appearance of the upward electric field $E_z$ before moderate and strong earthquakes indicates that the daily normal positive atmospheric electric field $E_z$ and the part induced by space weather (*14*) had been overcome. Therefore, $E_z$ presents some negative signatures, which may last for several hours or more. Radon gas ionizing the air after its escape from the rocks due to strong tectonic activity near the epicenter region has been suggested by previous researchers (*11, 15-21*). The physical process that produces an hourly scale persistent negative anomaly of the atmospheric electrostatic field near the epicenter before the earthquake is illustrated in Fig. 2.

Figure 2 shows the physical mechanism by which the seismogenic process changes the direction and magnitude of $E_z$. Fig. 2A: In a fair weather background, the normal vertical atmospheric electric field $E_z$ orients downward (the downward direction of $E_z$ is defined as positive under fair weather conditions). The seismogenic process always releases radon from many rock fractures. In turn, radon could generate radioactive decay and emit α, β and γ. These rays may further ionize the air nearby. With increasing emission of radon gases, many positively and negatively charged particles are injected close to the surface near the epicenter. Fig. 2B: Greater numbers of electron-ion pairs disperse in the air above the local region near the epicenter. Fig. 2C: Positive ions and negative ions, which bring positive charges and negative charges, are separated by the fair weather electric static field $E_z$, gravity and thermal convection, and move down and up, respectively. Fig. 2D: The earthquake-related $E_z$ (upward direction: negative) represents the electric field generated by a series of physical processes in which radon ionizes neutral gas in the air. This illustrates that in the last stage before the earthquake, the large release of radon gas near the imminent fracture region finally leads to the vertical atmospheric electrostatic field $E_z$ showing a unique negative (upward) orientation.

As shown in Fig. 2, the positively charged particles emitted at various altitudes are always driven downward by the fair weather $E_z$. Negatively charged particles remain at higher altitudes. Therefore, the gradual separation and accumulation of positive and negative ions produce an upward $E_z$. In other words, a reverse $E_z$ is formed near the Earth's surface. Finally, a very strong and stable earthquake-related electrostatic field is established and shows abnormal negative $E_z$.

It takes only several minutes to make the air more ionized from radon undergoing radioactive decay, while the time scale from the beginning of the abnormal $E_z$ signal to the EQ shock is approximately several hours to one day. It is demonstrated that there is a special process in which the underground rock structure experiences high pressures, and a large number of microfractures are formed. The interface between faults does not face so-called friction. Once a stress change arrives, it triggers an EQ to occur. Therefore, once an abnormal signal appears, it means that the whole fault is destabilizing. Observational experience tells us that the duration from this time until an EQ occurrence is only approximately several hours to one day.

Thus, geological movement determines that the abnormal $E_z$ signal is a very short-term precursor to an imminent emergency.

### Determining the Epicenter and EQ Magnitude by an $E_z$ Monitoring Network

Figure 3 is a cartoon map that indicates that a local monitoring network could be useful in warning for major earthquakes.

Figure 1A also illustrates that a pre-earthquake electrostatic field appeared in the Wenchuan region and nearby areas immediately before the Ms 8.0 earthquake. The earthquake-related reverse electrostatic field may be inhomogeneous over the whole fault region. The closer to the epicenter, the greater the negative $E_z$ value measured by the instrument. Therefore, before the Wenchuan earthquake, the value of -$E_z$ measured at Pixian station (50 km from the epicenter) was greater (4 times) than that measured at Wenjiang station (55 km from the epicenter), while both stations simultaneously showed that -$E_z$ reversals appeared from 7:00 LT until the earthquake occurred. The negative $E_z$ anomaly lasted for 7 hours (as shown in Fig. 1A). The comparison between the two simultaneous $E_z$ values from different observation sites demonstrates that the closer the $E_z$ monitoring station is to the epicenter, the greater the magnitude of the earthquake-related reverse $E_z$ signal obtained by the monitoring site. Based on the above observations, monitoring stations that may observe the reverse $E_z$ signal should be less than 100 km from the epicenter. If there are many nearby stations, someone could determine the location of the epicenter by $E_z$ data from multiple stations and triangulations.

It is suggested that with the establishment of a dense network of stations to monitor $E_z$ at more stations in different locations, one can more accurately determine the magnitude, timing, and location of major earthquakes. Geological conditions, the underlying surface structure in the epicentral region, and fair-weather wind speed and wind direction near the station during observation affect the magnitude and duration of the negative $E_z$ values related to the EQ. If a strong land EQ is forthcoming, the weather in the area approximately 50 km from the epicenter must be fair from several hours to one day immediately before the EQ (based on another paper by the authors). Therefore, the selection of the observation sites should be established according to the localized patterns of the fault zones in the future.

The stronger an earthquake is, the greater the size of the EQ seismogenic zone, and the more stations needed to detect negative $E_z$ values. According to the formula $R=10^{0.43M}+Cr$ (coefficient revised based on local conditions) (*22*), the magnitude of a potential earthquake could be proportional to the spatial scale of the fault. The monitoring system can calibrate some warning parameters based on the negative magnitude of the $E_z$ according to localized geological conditions. Once in fair weather conditions, when a station obtains an obvious stable abnormal negative $E_z$ value, a forthcoming earthquake might occur nearby in several hours to one day.

The method to determine a forthcoming EQ's epicenter location and the EQ magnitude is illustrated in a cartoon figure. As shown in Fig. 3, there are stations along a potential fault. The triangles denote stations. The red circle denotes the epicenter. $-E_z$ denotes a station that has observed abnormal negative atmospheric electrostatic signals under fair weather conditions. Fig. 3A illustrates that for an Ms 5 earthquake, anomalous negative $E_z$ signals may be recorded by 3 stations that may occupy a 100 km-long region by $R=10^{0.43M}+Cr$. Fig. 3B illustrates that for an Ms 6 earthquake, anomalous negative $E_z$ signals may be recorded by 7 stations that may occupy a 200 km-long region. Fig. 3C illustrates that for an Ms 7 earthquake, anomalous negative $E_z$ signals may be recorded by 15 stations that may occupy an 800 km-long region. Fig. 3D illustrates that for an Ms 8 earthquake, anomalous negative $E_z$ signals may be recorded by 21 stations that may occupy a 1500 km-long region.

**Summary**

1) Many observations, including signal features and an EQ list, have demonstrated that fair weather anomalous negative $E_z$ signals always appear several hours to one day before a forthcoming strong land earthquake.

2) During approximately the last several hours to one day before the strong land EQ, heat radiation makes the weather fair in the region near the epicenter and associated fault. Moreover, it is critical that a great number of microfractures are formed, and passages from deep rocks to the air are open so that rock fractures are connected. Underground gases, including radioactive matter, are released to the air with vapor. Air ionization is enhanced by many orders of magnitude. Positive and negative charges are separated by the electrostatic field, gravity and thermal convection. Finally, an additional atmospheric electrostatic field is generated and oriented against the fair weather atmospheric electrostatic field direction, leading to an obvious negative anomaly in the $E_z$ signal, which can easily be recognized by automatic software.

3) An available method is proposed in which by establishing a monitoring system that can detect the atmospheric electrostatic field, one can determine the epicenter and magnitude of a forthcoming strong earthquake.

4) The hourly warning concept might make all humans exempt from strong land earthquake hazards in the near future.

Therefore, the fair weather anomalous negative $E_z$ could be used as an important imminent (from approximately several hours to one day before the EQ) precursor of major earthquakes in a network monitoring system. Fig. 3 illustrates that by a special atmospheric electrostatic field monitoring network, an automatic system could provide information on the epicenter and the magnitude of a strong land EQ very quickly. Therefore, hourly warning for strong land earthquakes is possible!

**Acknowledgments:** The authors thank Prof. Fushan Luo, Prof. Jie Liu, Prof. Zhijun Niu and Prof. Shi Che for very useful discussions.

**Funding:** The authors were supported by the Strategic Pioneer Program on Space Science, Chinese Academy of Sciences, Grant No. XDA17010301, XDA17040505, XDA15052500, and XDA15350201, and by the National Natural Science Foundation of China, Grant No. 41874176 and 41931073. The authors thank the Chinese Meridian Project, the Specialized Research Fund for State Key Laboratories and CAS-NSSC-135 project (Grant No. Y92111BA8S), and the NSSC director fund (Grant No. E0PD41A11S).

**Author contributions:** Tao Chen designed the project and wrote the paper. Han Wu, Xiao-Xin Zhang, Xiao-Bing Jin, Lei Li, Shuo Ti, Jian-Guang Guo, and Cong Wang collected the atmospheric electrostatic field $E_z$ data and analyzed them compared with the earthquake data. Professors Chi Wang, Ji-Yao Xu, Hui Li, Yong Liu, Su-Ping Duan, Sai-Guan Xiao, Lei Dai, Ilan Roth, Vladimir Makhumutov and Xiong Hu contributed to space physics, atmospheric physics, and atomic physics theory and signal analysis. Qi-Ming Ma, Jia-Jun Song, Fang Xiao, Jing Luo, Zhao-Hai He, Chen Zeng, and Xiu-Jie Jiang were responsible for building the atmospheric electric field instrument, calibration work and the $E_z$ measurements. Xi-Chen Wang, Xu-Hui Shen, Shi Che, Xue-Min Zhang, Zhima Zeren, Fu-Qiong Huang, Jian-Ping Huang, Jing Liu, Pan Xiong, Quan Guo, Xiao-Dong Peng, Han-Yin Cui, Chao Li, Qiang Sun, Li Yao, Ai-Min Du, Yong Wei, Fei He, Hong Yuan, Sheng Wang, Yi Liu, Jing Yang, Hai-Bin Li, Jian-Feng Su, Wen Li, Da-Lin Li, Shi-Han Wang, Zhao-Hui Huang, Ling-Qian Zhang, Yong-Cun Zhang, Chun-Lin Cai, and Guang-Qing Yan provided subsonic and earthquake data and performed some relative data analysis. Tao Chen

**Competing interests:** Authors declare that they have no competing interests.

**Data and materials availability:** All data are provided by the China Meteorological Administration, China Earthquake Administration and Chinese Academy of Sciences.

**Tables:**

**Table 1. The strong land earthquakes recorded in references.**

| Date | Time (UTC) | Magnitude | Epicenter | Depth | Advance Time | VEF Depth | VEF Duration | Distance from Epicenter | Citation | Type | Weather |
|---|---|---|---|---|---|---|---|---|---|---|---|
| yyyy/mm/dd | hh:mm:ss | (M) | ° | (km) | (h) | (-kV/m) | (h) | (km) | | 1: Remains negative; 2: Returns to positive | |
| 1976/08/23 | 03:30 | 7.2 | 32.5N, 104.3E; China | 23 | 23 | 17 | 26.5 | 50 | Hao et al. (*5*) | 1 | Fair |
| 1986/08/30 | 21:28 | 6.9 | Poland | -- | 18.5 | 0.28 | 2 | 700 | Nikiforova and Michnowski (*23*) | 2 | Fair |
| 1992/03/05 | 14:39 | 6.4 | 52.9N, 159.62E; Russia | 45 | 9.5 | 0.4 | 1 | 130 | Rulenko et al. (*24*) | 2 | Fair |
| 1999/09/18 | 21:28 | 6.0 | 51.21N, 157.56E; Russia | 60 | 29.5 | 0.5 | 8 | 200 | Mikhailova et al. (*25*); Smirnov et al. (*26*) and Mikhailova et al. (*27*) | 2 | Fair |
| 2002/10/16 | 10:12 | 6.2 | 51.95N, 157.32E; Russia | 102 | 34 | 0.25 | 1 | -- | Mikhailov et al. (*28*) | 2 | Fair |
| 2008/05/12 | 06:28 | 8.0 | 30.95N, 103.4E; China | 33 | 7 | 2.7/0.7 | 7 | 50/55 | Chen et al. (*29*) | 1 | Fair |
| 2010/06/12 | 19:26 | 7.8 | 7.88N, 91.94E; India | 10 | 14 | 1.39 | 0.67 | 1763 | Choudhury et al. (*4*) | 2 | Fair |
| 2013/01/28 | 16:38 | 6.1 | 42.64N, 79.76E; Russia | 10 | 20 | 2 | 0.5 | 264 | Antonova and Zhumabaev (*30*) | 2 | Fair |
| 2019/06/17 | 14:55 | 6.0 | 28.34N, 104.9E; China | 16 | 23 | 11 | 1 | 59 | Chen et al. (*29*) | 2 | Fair |

**Figures:**

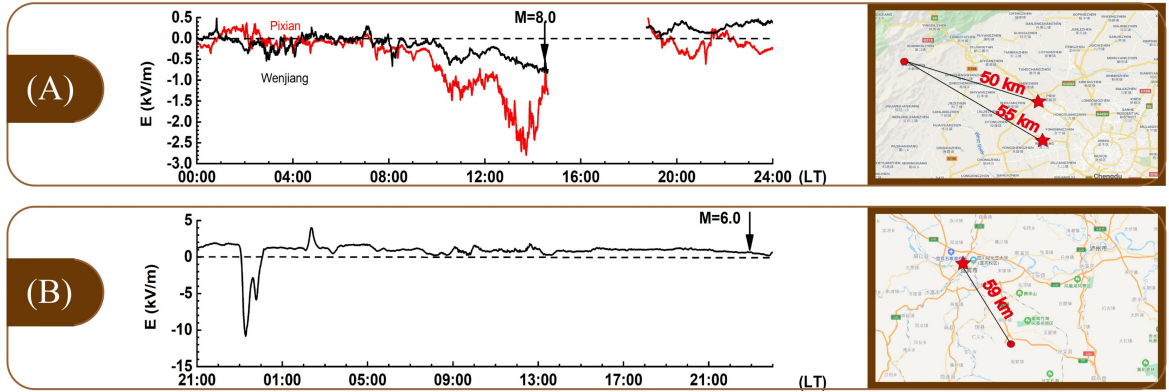

**Fig. 1. Negative atmospheric electrostatic field E$_z$ before earthquakes.** (**A**) Wenchuan M8.0 earthquake on May 12, 2008, with two different stations 50 and 55 km from the epicenter. (**B**) Changning M6.0 earthquake on June 17, 2019, with the station 50 km from the epicenter. The data gap in panel (B) is due to a power outage. In the maps on the right, the stars denote the stations, and the circles denote the epicenters.

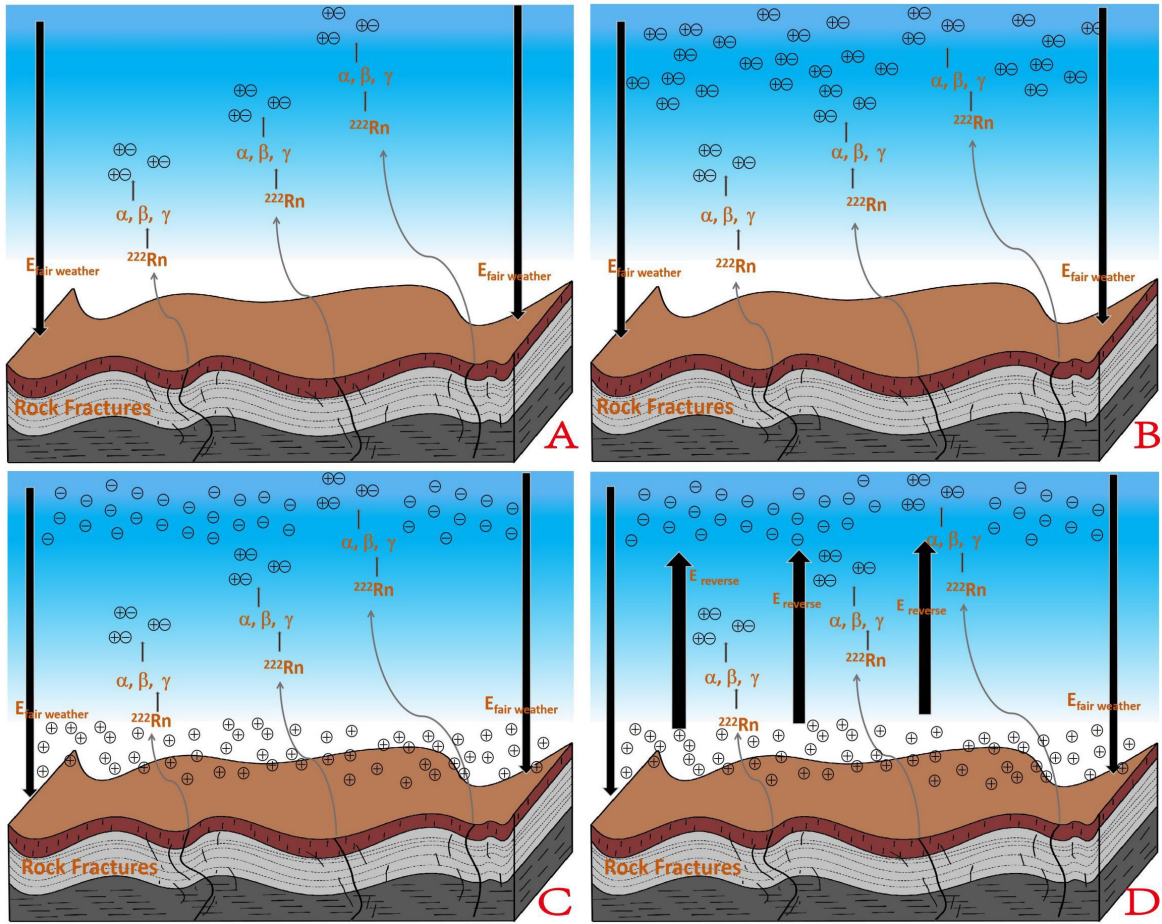

**Fig. 2. Schematic diagram of the seismogenic process changing $E_z$.** (**A**) In a fair weather background, the normal vertical atmospheric electric field $E_z$ orients downward (the downward direction of $E_z$ is defined as positive under fair weather conditions). The seismogenic process releases radon from rock fractures and produces α particle decay to complete ionizing radiation, so many positively and negatively charged particles are injected close to the surface near the epicenter. (**B**) Greater numbers of electron-ion pairs disperse in the air above the local region near the epicenter. (**C**) Positive ions and negative ions, which bring different charges, are separated by the fair weather electric static field $E_z$, gravity and thermal convection, and move down and up, respectively. (**D**) The earthquake-related $E_z$ (upward direction: negative) represents

the electric field generated by a series of physical processes in which radon ionizes neutral gas in the air.

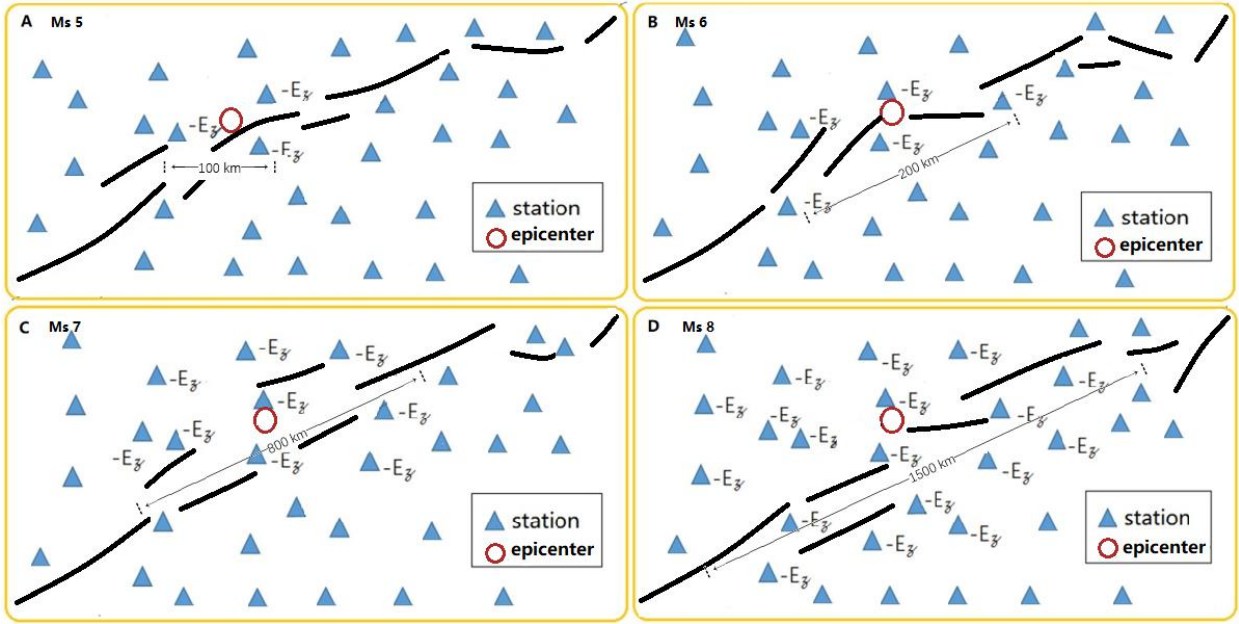

**Fig. 3. Cartoon map of the monitoring network for strong earthquake warning**. There are stations along a potential fault. The triangles denote stations. The red circle denotes the epicenter. $-E_z$ denotes a station that has observed abnormal negative atmospheric electrostatic signals. (**A**) For Ms 5, anomalous negative $E_z$ signals may be recorded by 3 stations that may occupy a 100 km-long region by R=$10^{0.43M}$-Cr (local factor). (**B**) For Ms 6, anomalous negative $E_z$ signals may be recorded by 7 stations that may occupy a 200 km-long region. (**C**) For Ms 7, anomalous negative $E_z$ signals may be recorded by 15 stations that may occupy an 800 km-long region. (**D**) For Ms 8, anomalous negative $E_z$ signals may be recorded by 21 stations that may occupy a 1500 km-long region.